\documentclass[11pt]{article}
\usepackage{amssymb}
\usepackage{mathtools}
\usepackage{epsf}
\usepackage{caption}
\usepackage{amsmath}
\usepackage{amsthm}

\newcommand{\oR}{{\mathbb R}}
\newcommand{\oE}{{\mathbb E}}

\newcommand{\oN}{{\mathbb N}}
\newcommand{\nn}{{\bf n}}
\newcommand{\ee}{{\bf e}}
\newcommand{\var}{{\rm{Var}}}
\newcommand{\cov}{{\rm{Cov}}}

\newtheorem{algo}{Algorithm}
\newtheorem{example}{Example}

\begin{document}
\thispagestyle{empty}

\begin{center}
{\bf \large A Cox rate-and-state model for monitoring seismic hazard \\
in the Groningen gas field}\\[0.5in]

Z. Baki $\mbox{}^{a}$ and
M.N.M. van Lieshout $\mbox{}^{b,a}$\\[0.2in]

$\mbox{}^a$
Faculty of Electrical Engineering, Mathematics and Computer Science,  \\
University of Twente, P.O.~Box 217, NL-7500 AE, Enschede, The Netherlands.

\

$\mbox{}^b$
CWI, P.O.~Box 94079, NL-1090 GB Amsterdam, The Netherlands 
\end{center}

\bigskip

\noindent
{\bf Abstract:} To monitor the seismic hazard in the Groningen gas field,
we modify the rate-and-state model that relates changes in pore
pressure to induced seismic hazard by allowing for noise in pore pressure 
measurements and by explicitly taking into account gas production volumes.
We analyse the first and second-moment structure of the resulting Cox
process, propose an unbiased estimating equation approach for the 
unknown model parameters and derive the posterior distribution of the 
driving random measure. We use a parallel Metropolis adjusted Langevin 
algorithm for sampling from the posterior and to monitor the hazard. 

\bigskip

\noindent
{\bf Keywords:} Cox process, gas production, induced seismicity, pore pressure,
rate-and-state model, spatio-temporal point process.

\bigskip

\noindent
{\bf Mathematics Subject Classification (MSC 2020):} 60G55, 62F15, 62M30.

\section{Introduction}
\label{S:intro}

The study of induced earthquakes caused by extraction or injection 
of fluids or gases is an important research topic. In the Netherlands, 
the Groningen gas field discovered in the late 1950s has played an 
important role in the Dutch economy. With an estimated recoverable
gas volume of over 2,900 billion Normal cubic metres spread over a 
region of about 900 square kilometres, it is one of the largest 
gas fields on the planet \cite{JageViss17}. However, large production
volumes in the 1970s caused a drop in pore pressure in the gas field
which resulted in induced earthquakes in the previously seismically
inactive region. Thus, it is essential to be able to predict seismic 
hazard based on field measurements, for instance of pore pressure or, 
equivalently, Coulomb stress. One of the most widely used methodologies 
to do so is the rate-and-state model \cite{Cand19, DempSuck17, Rich20} 
which is now considered to be the state-of-the-art technique \cite{Kuhn22}.

In the rate-and-state model (e.g.\ Candela et al. \cite{Cand19},
Dempsey and Suckale \cite{DempSuck17} and Richter et al. \cite{Rich20}),
the earthquakes follow a Poisson point process whose intensity function
$\lambda$ (the rate) is assumed to be inversely proportional to a state 
variable $\Gamma$ that is defined by an ordinary differential equation. 
This differential equation is based on physical considerations and takes 
into account the elapsed time and the change in pore pressure. Nevertheless,
it can be criticised on several points. Firstly, since, by definition,
the points in any Poisson point process do not interact with one another, 
the model is unable to deal with clustering as seen in, for instance,
the Groningen data \cite{LiesBaki24}. Secondly, the pressure values are 
assumed to be known everywhere. Lastly, the varying gas extraction is not 
taken into account. Our goal is to propose stochastic rate-and-state models 
that address these shortcomings and to develop a toolbox for statistical 
inference.

The plan is as follows. First, we briefly describe the data at our disposal 
and review the Poisson rate-and-state model. In Section~\ref{S:Cox}, we 
propose a modified Cox rate-and-state model. We give explicit expressions
of the first and second moments of the state variable in Section~\ref{S:state} 
and apply the delta method to approximate the first two moments of the rate 
variable. Some illustrations of the accuracy of the approximations for 
various pore pressure scenarios are given too (cf.\ Section~\ref{S:approxRate}). 
In Section~\ref{S:params}, 
we turn to the estimation of the model parameters. Next, Section~\ref{S:monitor} 
focuses on the random state variable. We calculate its posterior distribution 
given earthquake count data and study a parallel Metropolis adjusted Langevin 
monitoring algorithm, which is next applied to the Groningen data. We close 
with a discussion and some suggestions for future research.

\section{The Groningen gas field}

\subsection{Data}
\label{S:data}

\begin{figure}[thb]
\centering
\centerline{
\epsfxsize=0.45\hsize
\epsffile{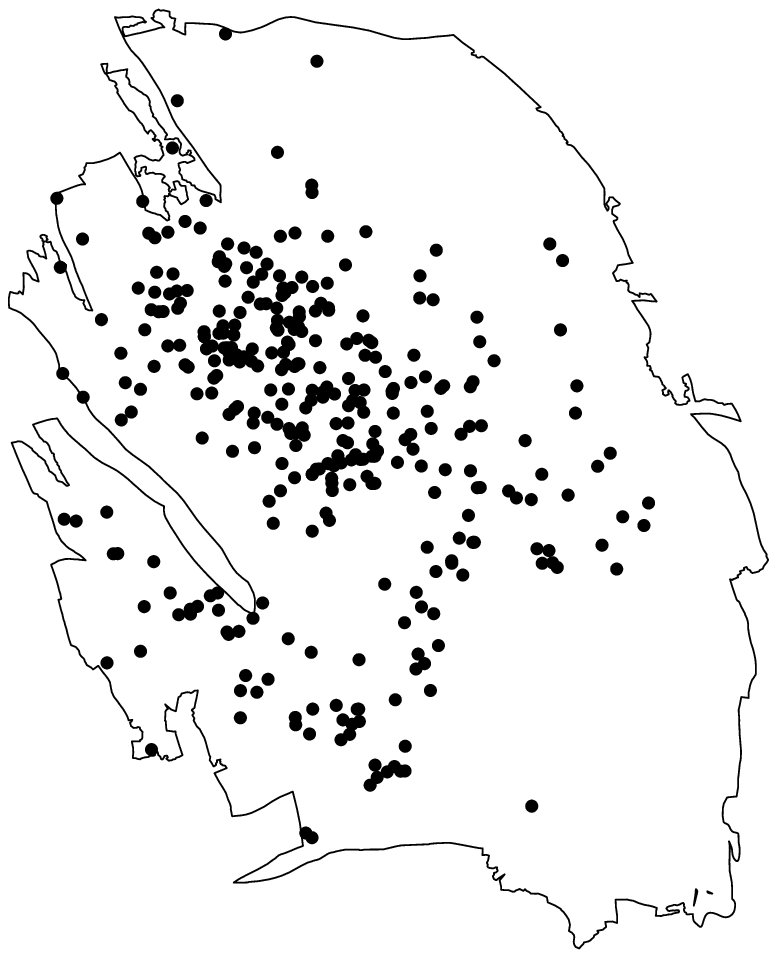}
\epsfxsize=0.45\hsize
\epsffile{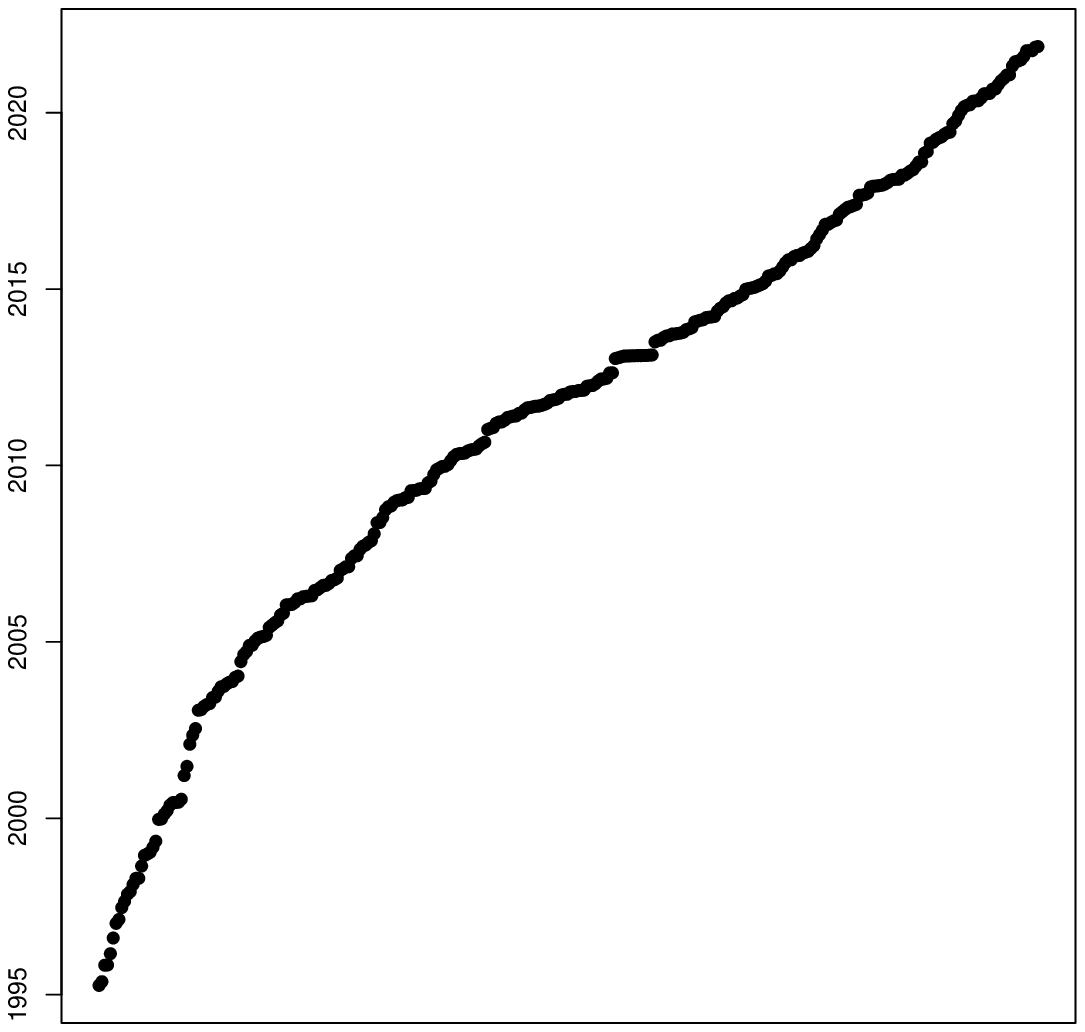}
}
\caption{Spatial (left-most panel) and temporal (right-most panel) 
projections of the $332$ earthquakes of magnitude $1.5$ or larger with 
epicentre in the Groningen gas field that occurred in the period from 
January 1st, 1995, up to December 31st, 2021.}
\label{F:eqs}
\end{figure}

Figure~\ref{F:eqs} shows the spatial and temporal projections of
the $332$ earthquakes of magnitude $1.5$ or larger that occurred from 
January 1st, 1995, up to December 31st, 2021, in the Groningen gas
field\footnote{Catalogue from {\tt{www.knmi.nl}}, shapefiles 
from {\tt{www.nlog.nl}} downloaded April 2022}. Note that earthquakes
seem to happen more often in the central and southwestern parts 
of the gas field. Temporally, the steeper curve in the 1990s reflects 
the longer spell between successive earthquake occurrences; flatter 
pieces indicate a quicker succession of earthquakes. 

To explain the observed heterogeneity, two covariates are at our
disposal. Monthly production values from the start of preliminary 
exploration in February 1956 up to and including December 2021
were kindly provided to us\footnote{Mr Rob van Eijs from Shell}.
Additionally, $2009$ pore pressure observations are available over 
the period from April 1960 until November 2018\footnote{{\tt 
nam-feitenencijfers.data-app.html/gasdruk.html}}. For a fuller 
discussion, we refer to \cite{LiesBaki24}.

\subsection{The Poisson rate-and-state model}
\label{S:Poisson}

In the classic rate-and-state model \cite{Cand19,DempSuck17}, 
earthquakes occur according to a spatio-temporal Poisson point 
process with intensity function 
\begin{equation}
\lambda(s,t) = r_0 \frac{\Gamma(s,0)}{\Gamma(s,t)}, \quad
(s,t) \in W_S \times W_T,
\label{e:PoisIntens}
\end{equation}
where $W_S \subset \oR^2$ is a compact subset of the plane 
and $W_T$ a closed and bounded interval in $\oR$. The parameter
$r_0 > 0$ is the background seismicity, the state variable 
$\Gamma(s,t)$ is defined by the ordinary differential equation
\[
d\Gamma(s,t) = \alpha \left[ dt + \Gamma(s,t) dX(s,t) \right],
\]
where $X(s,t)$ is the pore pressure at spatial location $s$ and time $t$
and $\alpha > 0$. Multiplying both sides by $\exp( -\alpha X(s,t) )$,
discretising in time steps of length $\Delta > 0$ and writing $t_k = 
t_0 + k\Delta$ for the $k$-th point from $t_0 = \min \{t: t\in W_T \}$, 
$k=1, \dots, m$, we obtain the Euler difference equation 
\begin{equation}
\Gamma(s, t_{k +1} ) = \left( 
    \Gamma(s, t_k) + \alpha \Delta \right) 
\exp\left[ \alpha ( X(s,t_{k+1} )  - X(s,t_k) ) \right],
\quad s \in W_S.
\label{e:Euler}
\end{equation}
The parameters $r_0$, $\alpha$ and the initial state $\Gamma(s, t_0) 
\equiv \gamma_0 > 0$ are treated as unknowns and can be estimated, for 
example, by the maximum likelihood method. For a full discussion and
comparison with other techniques, we refer to \cite{Kuhn22}. In line 
with current practice, we also discretise the spatial domain $W_S$ in 
a regular grid with cell representatives  $s_1, \dots, s_n$. Note that
because $W_S$ is not necessarily rectangular, grid cells may have 
different areas, which we shall denote by $\Delta(s_i)$.

\section{The Cox rate-and-state model}
\label{S:Cox}

The Poisson rate-and-state model of Section~\ref{S:Poisson} expresses
the change in seismic hazard in terms of elapsed time and pore pressure
change. In practice, the values of the latter are typically available
only at wells and monitoring stations. At other locations, the pore
pressure must be estimated \cite{LiesBaki24} or approximated by linear 
(or spline-based) interpolation \cite{Rich20}. For the Groningen data
described in Section~\ref{S:data}, there are only some two thousand 
observations scattered unevenly over the field and spanning a period 
of more than sixty years. Therefore, it would be better to explicitly 
take the uncertainty into account and model the $X(s_i, t_j)$ as a 
random variable. By doing so, we obtain a doubly stochastic or Cox point 
process \cite{SKM}. Briefly, given a realisation $\lambda$ of the density
$\Lambda$ of a random measure on $W_S\times W_T$, the driving random 
measure of the Cox process, the earthquakes form a Poisson process with 
intensity function $\lambda$. Thus, the distribution of the Cox process is 
fully characterised by the distribution of the driving random measure. 

To find a suitable driving random measure, we assume that the pore pressure 
can be decomposed in a deterministic and stochastic part, that is,
\begin{equation}
X(s_i, t_j) = c(s_i, t_j) ^\top \beta + E(s_i, t_j)
\label{e:X}
\end{equation}
for some known function $c$ with values in $\oR^p$, unknown
parameter $\beta \in \oR^p$ and independent mean-zero random variables
$E(s_i, t_j)$ with variance $\sigma^2$. The parameters $\beta$ and 
$\sigma^2$ can then be estimated by the least squares method. For our 
data, as in \cite{LiesBaki24}, without recourse to further explanatory 
variables, we let $c$ be a polynomial of order four in space, of order 
two in time and add interaction terms up to first order in time and third 
order in space.  For brevity, we will write 
\(
m(s_i, t_j) = c(s_i, t_j)^\top \hat \beta
\)
from now on. The variogram of the residuals is flat in time. The  
residual spatial variation $U(s_i)$ is therefore a function of spatial 
location only, which, since 
the rate-state equation depends only on temporal changes in pressure, may 
be ignored.

The other explanatory variable at our disposal is the vector of monthly 
production figures at the production wells, which can be smoothed out over 
the field \cite{LiesBaki24}. We model their influence on the earthquake 
intensity function through the multiplier $r_0$ in (\ref{e:PoisIntens}).
Specifically, with $\Delta$ equal to one year, write $V(s_i, t_j)$ for 
the gas extracted in the cell around $s_i$ over the year preceding $t_j$ 
and replace the constant $r_0$ by
 $\exp\left[ \theta_1 + \theta_2 V(s_i, t_j) \right]$.

In summary, we obtain a Cox process $\Psi$ with driving random measure
defined by its density function 
\begin{equation} \label{e:driving}
\Lambda(s,t) = 
\exp\left[ 
 \theta_1 +  \theta_2 V(s,t) 
\right] \frac{ \gamma_0}{ \Gamma(s,t)}
\end{equation}
which we discretise for computational convenience. Specifically, write
$N(s_i,t_j)$ for the number of earthquakes in the cell of $(s_i, t_j)$.
Then, conditional on $\Lambda(s_i, t_j)$, $N(s_i, t_j)$ is Poisson 
distributed with rate parameter $\Lambda(s_i,t_j) \Delta(s_i) \Delta$
independently of the earthquake counts in other cells. The Euler difference 
equation (\ref{e:Euler}) can be solved explicitly, and we obtain 
\begin{equation}
\Gamma(s_i, t_j)  = 
  \exp\left[ \alpha X(s_i, t_j) \right]
\left\{
 \alpha \Delta \sum_{k=0}^{j-1} \exp\left[ - \alpha X(s_i, t_k) \right]
+
    \gamma_0 \exp\left[ - \alpha X(s_i, t_0)  \right]
\right\} .
\label{e:Gamma}
\end{equation}

The parameters can be interpreted as follows. For positive $\theta_2$,
an increase in production tends to increase the number of earthquakes;
the real-valued parameter $\theta_1$ is the intercept. Mathematically,
the model is well-defined for non-positive $\theta_2$ as well but 
does not make practical sense. As for the classic model, $\gamma_0>0$ 
is the initial state of the stochastic difference equation. When $\alpha$ 
is positive and the pore pressure remains constant, the state variable 
increases over time and the earthquake hazard decreases according to
Omori's law \cite{Utsu95}. When the pore pressure changes due to gas 
extraction or fluid injection, for positive $\alpha$, an increase in 
pore pressure due to fluid injection emphasises the temporal increase 
in the state variable and thus reduces the seismic hazard even more. A 
drop in pore pressure leads to a decrease in the state variable; its 
effect on the intensity of earthquakes depends on the combined effect of 
time and pressure.

\section{Moments of the state variable}
\label{S:state}

Since the randomness in the driving random measure (\ref{e:driving}) of our 
Cox process is induced by the state process $\Gamma$ through (\ref{e:X}),
we investigate its first and second-moment properties first. Let $\Gamma$ 
be defined by (\ref{e:Gamma}) for some $s$ and the set of $t_j$, $j=0,\dots, m$. 
Set $c = \exp\left(\alpha^2 \sigma^2\right)$ and define, for $i,j \in\oN_0$,
\(
f_{ij} = 
\exp\left[ \alpha \left\{ m(s, t_i) - m(s, t_j) \right\} \right].
\)
Then, for $k\in\oN$,
\begin{eqnarray*}
\oE \Gamma(s, t_k) & = & c \left( \alpha \Delta \sum_{i=0}^{k-1} f_{ki} +
   \gamma_0 f_{k0} \right),\\
\var \Gamma(s, t_k) & = & 
\alpha^2 \Delta^2 c^2 (c^2 - 1) \sum_{i=0}^{k-1} f_{ki}^2 
+ \alpha^2 \Delta^2 c^2 (c-1) \sum_{i=0}^{k-1} 
\sum_{i\neq j=0}^{k-1}  f_{ki} f_{kj}  \\
& + & 
2 \alpha \Delta \gamma_0 c^2 f_{k0}^2 
   \left( c^2 - 1 + (c-1) \sum_{i=1}^{k-1} f_{0i} \right)
 + \gamma_0^2 f_{k0}^2 c^2 (c^2 - 1) 
\end{eqnarray*}
and, for $0 < k < l$,
\begin{eqnarray}
\cov( \Gamma(s, t_k), \Gamma(s, t_l) ) & =  &
\nonumber
\alpha^2 \Delta^2 c^2 \sum_{i=0}^{k-1} 
\left[ f_{ki} f_{li} (c-1) - f_{li} \left( 1 - \frac{1}{c} \right) \right] \\
& + & ( 2 \alpha \Delta \gamma_0 + \gamma_0^2) c^2 f_{k0} f_{l0} ( c - 1)
- \alpha \Delta \gamma_0 c^2 f_{l0} \left( 1 - \frac{1}{c} \right).
\label{e:covsum}
\end{eqnarray}

When fluid is injected into a field, the pore pressure typically increases.
In this case, the state variables are positively correlated, that is, 
\(
\cov( \Gamma(s, t_k), \Gamma(s, t_l) )  \geq 0
\)
for all $k, l \in\oN_0$.

When the pore pressure decreases, for example due to gas extraction,
the picture is more varied.
If $ \alpha \sigma^2 > m(0)$,
then $\cov( \Gamma(s, t_k), \Gamma(s, t_l) ) \geq 0$ for 
all $k,l\in\oN_0$.
On the other hand, if  
\(
\alpha \sigma^2 < \min_{i\in \oN_0} \left\{ m(s, t_i) - m(s, t_{i+1}) \right\},
\)
the minimal drop in pressure in between observation epochs,
\[
\cov( \Gamma(s, t_k), \Gamma(t, t_l) )
 - (\alpha \Delta \gamma_0 + \gamma_0^2) c^2 (c-1) f_{k0} f_{l0} \leq 0
\]
for all $0 \leq k < l$.
The proofs of these statements can be found in Appendix~A.
The following examples are illuminating.

\begin{table}
\begin{center}
\begin{tabular}{|l|rrrrrrr|}
\hline
Years & & & & & & & \\
\hline
1995--2001 & 179.81 & 177.39 & 174.86 & 172.20 & 169.42 &
 166.50 & 163.48 \\
2002--2008 & 160.32 & 157.05 & 153.65 & 150.13 & 146.49 & 
 142.72 & 138.82 \\
2009--2015 & 134.81 & 130.68 & 126.43 & 122.04 & 117.53 & 
 112.91 & 108.16 \\
2016--2021 & 103.28 &  98.29 &  93.17 &  87.94 &  82.56 & 
  77.08 &  \\
\hline
\end{tabular}
\end{center}
\caption{Estimated pore pressure in bara on January 1st in the years
1995--2021 near the town of Slochteren, The Netherlands.}
\label{T:Slochteren}
\end{table}

\begin{example}
{\rm
Table~\ref{T:Slochteren} lists estimated pore pressure values 
near the town of Slochteren in the Groningen gas field in The 
Netherlands for January 1st, 1995--2021 \cite{LiesBaki24}.
The estimated standard deviation is $\hat \sigma = 7.17$. 

Note that the pore pressure values are decreasing due to 
gas extraction. Since the intensity of induced earthquakes was 
very low in 1995, $\gamma_0$ can be considered infinite. 
Equation (\ref{e:covsum}) then implies that the covariance matrix 
of the random vector $\Gamma(s, t_k)_k$ has positive entries only.
}
\end{example}

\begin{example}
{\rm
Next, let us suppose that -- in contrast to the previous example --
the initial seismicity is very high, i.e.\ $\gamma_0 = 0$.
Assume a linearly decreasing sequence of pore pressures 
$m(s, t_0) = 3$, $m(s, t_1) = 2 $ and $m(s, t_2) = 1 $ and set $\alpha = 1$. Then
the covariance matrix of the random vector $(\Gamma(s, t_1), \Gamma(s, t_2))$
is readily calculated. Indeed,
$\var\Gamma(s, t_1) = e^{-2} ( e^{4\sigma^2} - e^{2 \sigma^2})$ and 
$\var\Gamma(s, t_2) = (e^{-2} + e^{-4}) ( e^{4\sigma^2} - e^{2\sigma^2})
+ 2 e^{-3} (e^{3\sigma^2} - e^{2\sigma^2})$. As for the off-diagonal
entry,
\begin{eqnarray*}
\cov( \Gamma(s, t_1), \Gamma(s, t_2) ) & = & \cov\left(
   e^{-1} e^{E(s, t_1) - E(s, t_0)}, e^{-2} e^{E(s, t_2)-E(s, t_0)} 
+ e^{-1} e^{ E(s, t_2) - E(s, t_1) } \right) \\
& = & e^{2\sigma^2} \left\{
   e^{-3} ( e^{\sigma^2} - 1 ) - e^{-2} ( 1 - e^{-\sigma^2} ) 
\right\}
\end{eqnarray*}
is negative for $\sigma^2 < 1$ and positive for $\sigma^2 > 1$.
}
\end{example}

\section{Approximate moments of the rate variable}
\label{S:approxRate}

Recall that in the rate-and-state model defined in Section~\ref{S:Cox},
the rate of induced earthquakes is inversely proportional to the state.
Due to its form (\ref{e:Gamma}), the moments of 
${1} / {\Gamma(s, t_k)}$ are intractable, but we can use the delta
method \cite{Vaar98} to approximate them in terms of the tractable moments 
of $\Gamma(s, t_k)$. Indeed, for $k = 0, 1, \dots$,
\begin{equation}
 \oE \left[ \frac{1}{\Gamma(s, t_k)} \right]
\approx
\frac{1}{\oE \Gamma(s, t_k)} 
+ \frac{\var \left( \Gamma(s, t_k ) \right)}
{ (\oE \Gamma(s, t_k))^3} .
\label{e:mean}
\end{equation}
Note that the approximation of the expectation of $\Gamma(s, t_k)^{-1}$
is at least as large as its `plug-in estimator' $1/ \oE \Gamma(s, t_k)$.

The same approach can be used to obtain an approximation for the covariance.
First, note that for $k, l\in\oN_0$,
\begin{eqnarray*}
 \oE \left[ \frac{1}{\Gamma(s, t_k) \Gamma(s, t_l)} \right] & \approx &
 \frac{1}{\oE \left[ \Gamma(s, t_k) \right] \, \oE\left[ \Gamma(s, t_l) \right]}  
 + 
\frac{ \cov( \Gamma(s, t_k ), \Gamma(s, t_l ) ) }{
(\oE \Gamma(s, t_k))^2 (\oE \Gamma(s, t_l))^2} \\
& + &  
\frac{\var \Gamma(s, t_k )}{ \oE \Gamma(s, t_l ) (\oE \Gamma(s, t_k))^3}
+
\frac{\var \Gamma(s, t_l )} { \oE \Gamma(s, t_k ) (\oE \Gamma(s, t_l))^3}.
\end{eqnarray*}
Plugging in expression (\ref{e:mean}), we obtain
\begin{equation}
\var\left( \frac{1}{\Gamma(s, t_k)} \right)
\approx
\frac{
\var( \Gamma(s, t_k ) )
}{
(\oE \Gamma(s, t_k))^4}.
\label{e:cov}
\end{equation}
For details, see Appendix~B.

\begin{figure}
\begin{center}
\epsfxsize=0.4\hsize
\epsffile{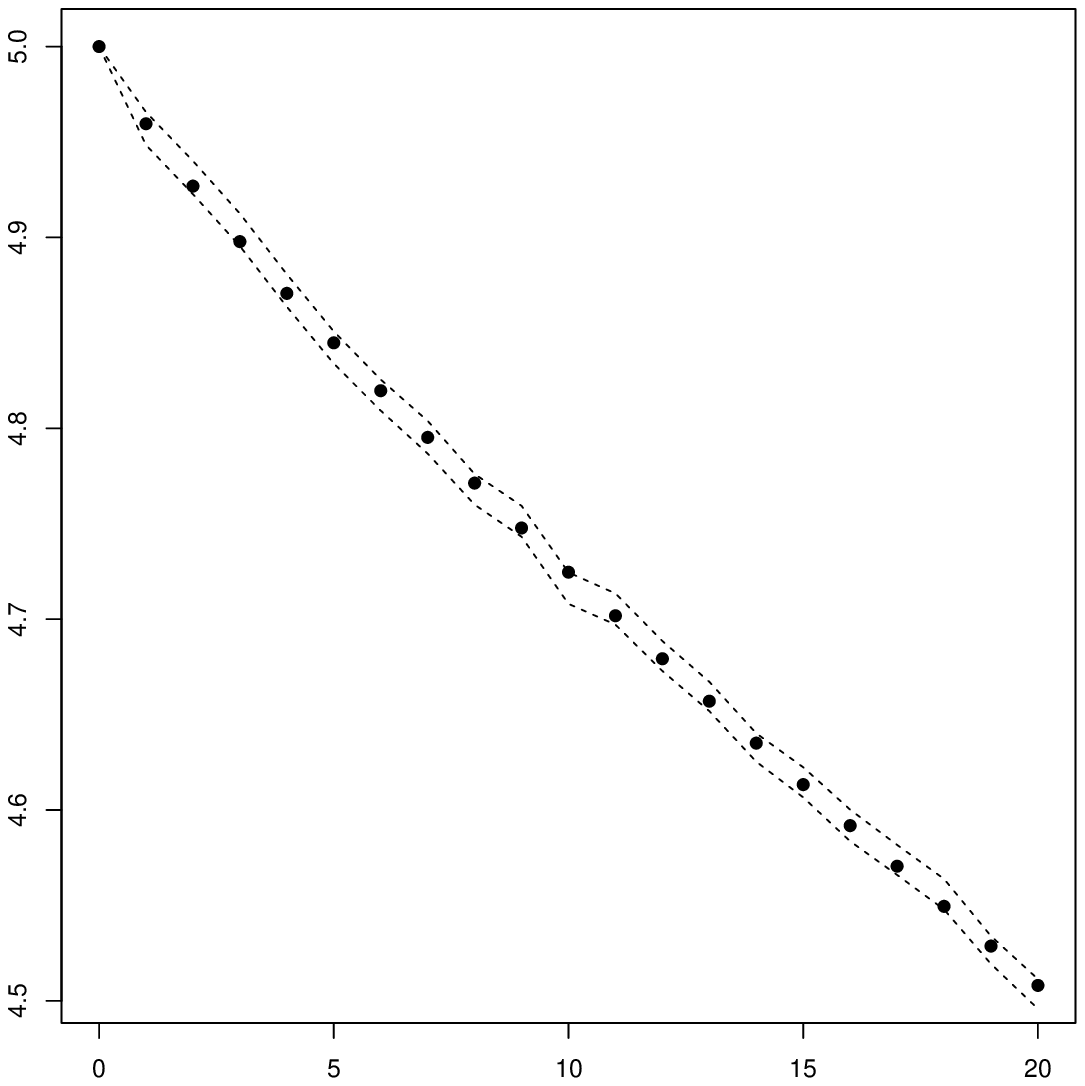}
\epsfxsize=0.4\hsize
\epsffile{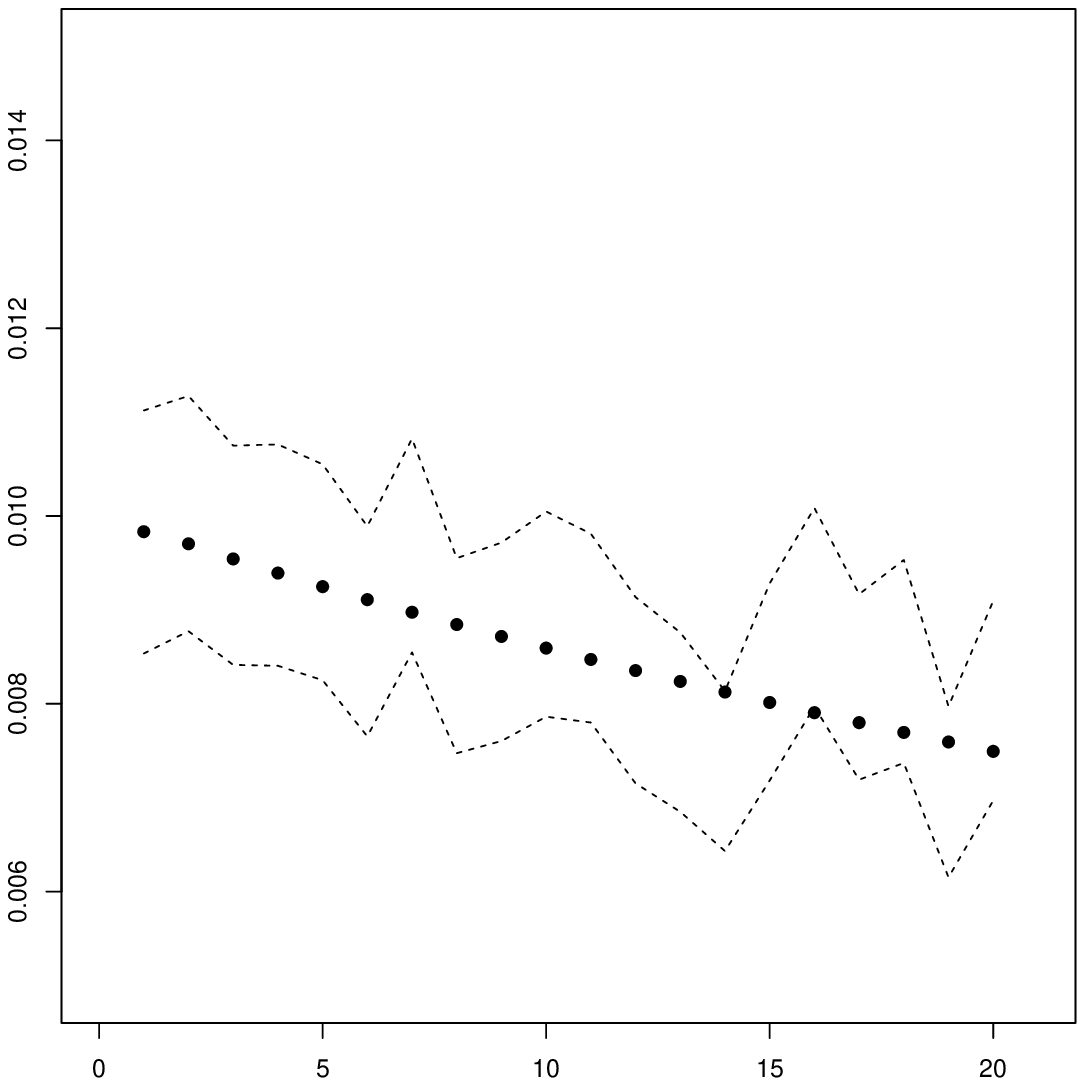}
\end{center}
\caption{$95\%$ pointwise confidence intervals for the mean (left) and variance (right) of
$\Gamma(s, t_k)^{-1}$ as a function of $k$ when $m(s, t_k ) = 6 - {1}/ { (0.5 k + 1) }$,
$\alpha = 0.01$, $\Delta = 0.1$, $\gamma_0 = 0.2$ and $\sigma^2 = 2.0$.
The dots correspond to the approximations in (\ref{e:mean}) and (\ref{e:cov}).}
\label{F:fig1ab}
\end{figure}

\begin{figure}
\begin{center}
\epsfxsize=0.4\hsize
\epsffile{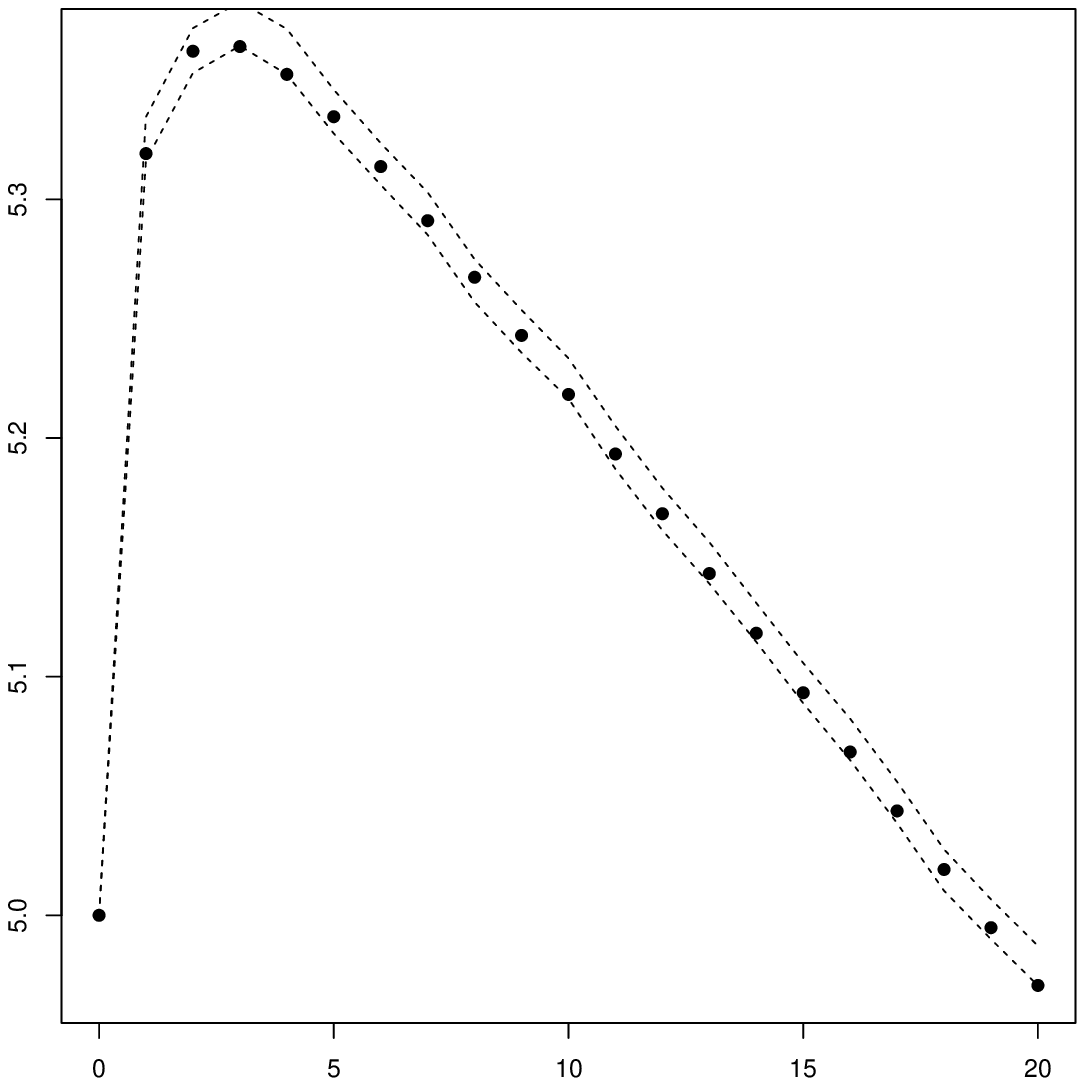}
\epsfxsize=0.4\hsize
\epsffile{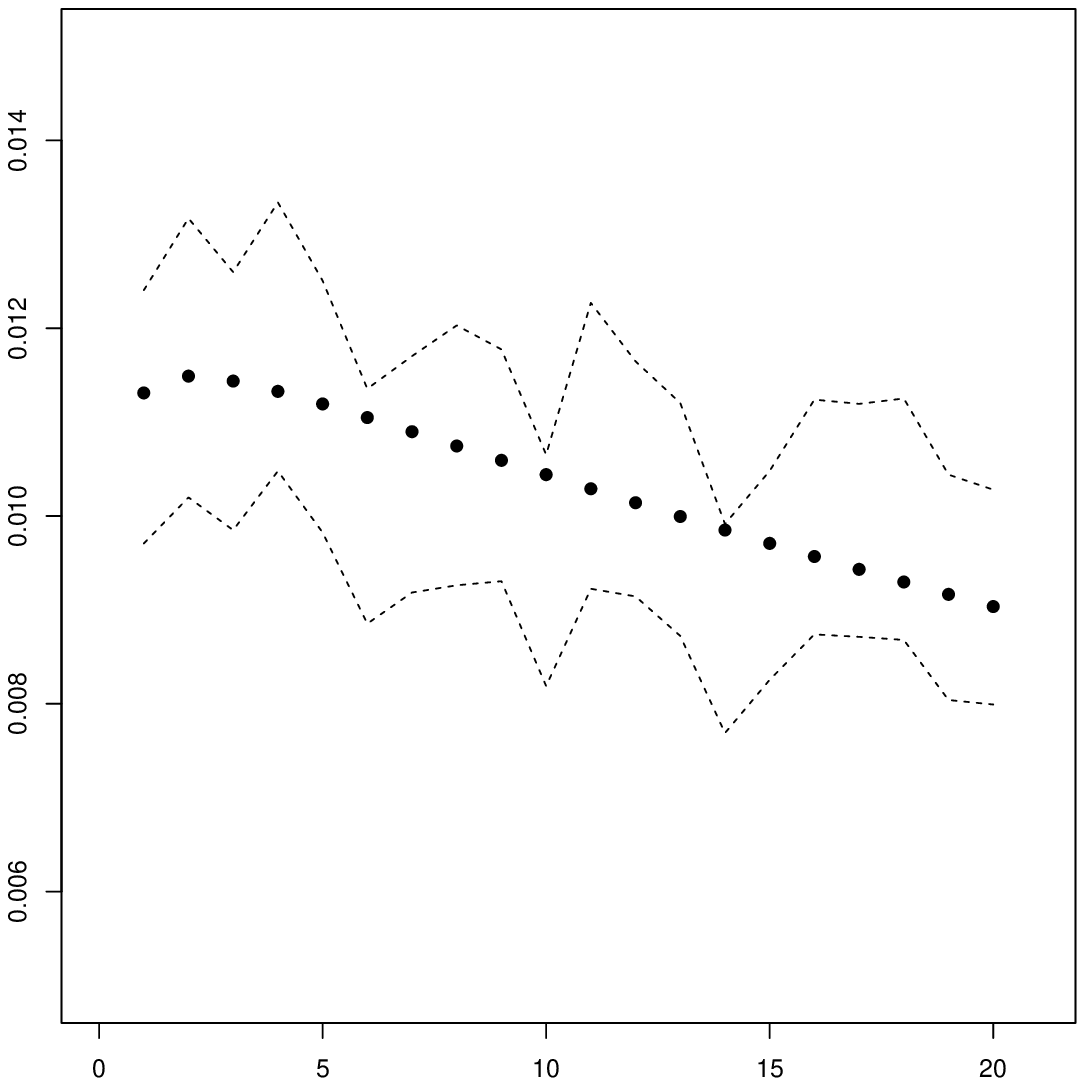}
\end{center}
\caption{$95\%$ pointwise confidence intervals for the mean (left) and variance (right) of
$\Gamma(s, t_k)^{-1}$ as a function of $k$ when $m(s, t_k) = 5 + {10} /  {( 2 k + 1) }$,
$\alpha = 0.01$, $\Delta = 0.1$, $\gamma_0 = 0.2$ and $\sigma^2 = 2.0$.
The dots correspond to the approximations in (\ref{e:mean}) and (\ref{e:cov}).}
\label{F:fig2ab}
\end{figure}

To investigate the accuracy of the approximations, we compare the approximation to population
estimates.  Recall that for an i.i.d.\ sample $Y_1, \dots, Y_n$ of some random variable
$Y_1$ with mean $\mu$ and variance $\sigma^2$,
approximate confidence intervals for $\mu$ and $\sigma^2$ take the form
\[
  \left( \bar Y_n - \frac{S_n}{\sqrt n} \xi_{1-\alpha/2}, \quad
         \bar Y_n + \frac{S_n}{\sqrt n} \xi_{1-\alpha/2} \right)
\]
for $\mu$,
and 
\[
  \left( \frac{S_n^2}{1 + \xi_{1-\alpha/2} \sqrt( \frac{2 \zeta}{n-1} ) }, \quad
         \frac{S_n^2}{1 - \xi_{1-\alpha/2} \sqrt( \frac{2 \zeta}{n-1} ) } \right)
\]
for $\sigma^2$. Here $\bar Y_n$ and $S_n^2$ are the sample mean and variance,
and $\xi_{1-\alpha/2}$ is the $(1-\alpha/2)$-quantile of the 
standard normal distribution. Furthermore, 
$\zeta = \frac{1}{2} ( \gamma_4 - 1 ) $, where $\gamma_4$ denotes
the ratio of the fourth central moment and the squared variance.
Since $\gamma_4$ is unknown, we estimate it by
\(
\widehat{\gamma_4} = { n \sum(Y_i - m)^4 } / { ( (n-1)^2 S_n^4 ) }
\)
where $m$ is the sample median \cite{Curt23}.
A similar approach may be taken for the covariance, although we do not pursue it here.

We consider two cases: increasing and decreasing pore pressure. 
For Figure~\ref{F:fig1ab}, define the pore pressure by the increasing function
\[
m(s, t_k ) = 6 - \frac{1}{ 1 + k/2 }.
\]
With parameter values $\alpha = 0.01$, $\Delta = 0.1$, $\gamma_0 = 0.2$ and
$\sigma^2 = 2.0$, the $95\%$ pointwise confidence intervals
for the mean and variance are given in the left- and right-most panels.
In both cases, the sample size was $n=500$. It can be seen that the
approximations are quite adequate. 

For Figure~\ref{F:fig2ab}, take the decreasing function
\[
m(s, t_k ) = 5 + \frac{10}{ 2 k + 1 }.
\]
With the same parameter values and sample size, the $95\%$ pointwise confidence intervals
for the mean and variance are given in the left- and right-most panels.
Again, the approximations are satisfactory.

\section{Parameter estimation}
\label{S:params}

The Cox model (\ref{e:driving}) depends on several parameters:
$\theta_1$, $\theta_2$, $\sigma^2$, $\alpha$ and $\gamma_0$.
The parameters $\gamma_0$ and $\theta_1$ are not identifiable.
Thus, we follow \cite{Rich20} and re-parametrise in terms of 
$\alpha / \gamma_0$. Recall that $\alpha$ and $\gamma_0$ are 
assumed to be positive. We therefore apply a logarithmic 
transformation and set $\eta = \log (\alpha / \gamma_0)$. 
The parameter $\sigma^2$ quantifies the uncertainty in the
pore pressure observations and may be estimated by a least
squares approach, and the vector of remaining parameters will
be denoted by $\zeta = (\theta_1, \theta_2, \alpha, \eta)$. 
Since the likelihood of a Cox process is intractable 
\cite{MollWaag04}, we use an estimating equations approach 
\cite{Vaar98} for $\zeta$. 

Our inspiration is the unbiased estimating equation from \cite{Waag07}
based on the gradient of the Poisson likelihood function. However, 
this method cannot be applied directly since it assumes that the 
intensity function $\lambda$ is known analytically. For our Cox 
process, though,  
\[
\lambda(s_i, t_j)  =  
\alpha e^{ -\eta + \theta_1 + \theta_2 V(s_i,t_j)} 
\oE\left[ \Gamma(s_i,t_j)^{-1} \right] 
\]
depends on the intractable expectation of the rate variable
(cf.\ (\ref{e:mean})). Therefore, consider the modified estimating 
equation
\begin{equation}
\label{e:Rasmus-modified}
F( \zeta ) = \sum_{(s_i, t_j)} \frac{\nabla h(s_i, t_j; \zeta)}
{h(s_i, t_j; \zeta)} 
   \left[ N(s_i, t_j) - \widehat {\lambda(s_i, t_j; \zeta)}
   \Delta \Delta(s_i) \right] = 0
\end{equation}
where 
\[
h(s_i, t_j; \zeta) = 
\frac{
e^{ \theta_1 + \theta_2 V(s_i, t_j ) }
e^{ - \alpha m(s_i, t_j)} 
 }{
 e^\eta \Delta \sum_{k=0}^{j-1} e^{ - \alpha m(s_i, t_k) }
+ e^{ - \alpha  m(s_i, t_0) }
} 
\]
and $\hat \lambda$ is an estimator for $\lambda$. Note that the function 
$h$ is equal to the intensity function when there is no noise 
(i.e.\ $\sigma^2 = 0$), in which case (\ref{e:Rasmus-modified}) reduces
to Poisson likelihood estimation \cite{Waag07}. To estimate the 
intensity function, we use
\[
\widehat{\lambda(s_i, t_j; \zeta )} = 
e^{ \theta_1 + \theta_2 V(s_i, t_j ) }
\frac{1}{L} \sum_{l=1}^L
\frac{ e^{ - \alpha X_l(s_i, t_j )} 
 }{
 e^\eta \Delta \sum_{k=0}^{j-1} e^{ - \alpha X_l(s_i, t_k) }
+ e^{ - \alpha  X_l(s_i, t_0) }
} 
\]
over an independent sample $X_l$, $l=1, \dots, L$, of $X = m + E$. Since
\[
\oE_\zeta \left[ 
  N(s_i, t_j) - \widehat{\lambda(s_i, t_j; \zeta )} \Delta\Delta(s_i) \right] = 
\lambda(s_i,t_j; \zeta) \Delta \Delta(s_i) - 
\lambda(s_i,t_j; \zeta) \Delta \Delta(s_i) = 0, 
\]
(\ref{e:Rasmus-modified}) is unbiased. Equation (\ref{e:Rasmus-modified}) can be solved numerically. 

Returning to the Groningen data (cf.\ Section~\ref{S:data}), we discretise time in years,
with $t_0$ equal to January 1st, 1995, and space in a $32\times 32$ rectangular grid 
surrounding the gas field with $s_i$ the centres of the cells. From \cite{LiesBaki24}, 
$\hat \sigma = 7.17$. For the other parameters, $\eta$ is effectively $-\infty$,
$\theta_1=-5.3$, $\theta_2 = 9.7$ and $\alpha= 0.0097$ 
using $L=1,000$ samples of $X$ for $\hat \lambda$. 

The quality of an estimating equation is expressed in terms of the variance of $F(\zeta_0)$ under 
the true value $\zeta_0$ of the parameter vector. However, since multiplying the left- and 
right-hand side of (\ref{e:Rasmus-modified}) by the same constant does not alter the estimator 
but does affect the variance, one needs to fix the scale by $U = -\oE_{\zeta_0} J_F(\zeta_0)$, 
the expectation of the negative Jacobian \cite{Goda85,GodaHeyd10}. The variance of the scaled
estimating equation is known as the inverse Godambe matrix. We refer to Appendix~C for explicit 
expressions of $F(\zeta)$ and its Jacobian, to Appendix~D for an asymptotic expression of the Godambe
matrix when the discretisation mesh goes to zero.

Under an appropriate asymptotic scheme, for example by letting the discretisation get finer and 
finer, the observation window larger and larger, the inverse Godambe matrix can be interpreted
as the variance of $\hat \zeta$. However, for the Groningen data, in view of the small earthquake 
counts and numerical stability considerations, we discretise rather coarsely. Therefore it is better 
to use a parametric bootstrap approach \cite{Vaar98} to obtain approximate confidence intervals. 
This way, at $95\%$ confidence level, we obtained the confidence interval $(-5.6, -5.1)$ for $\theta_1$, 
$(7.0, 12.8)$ for $\theta_2$ and $(0.006, 0.01)$ for $\alpha$.
Note that the confidence interval for $\theta_2$ contains only strictly positive values,
indicating that an increase in production leads to a higher earthquake hazard the next year.

\begin{figure}[htb]
\centering
\centerline{
\epsfxsize=0.45\hsize
\epsffile{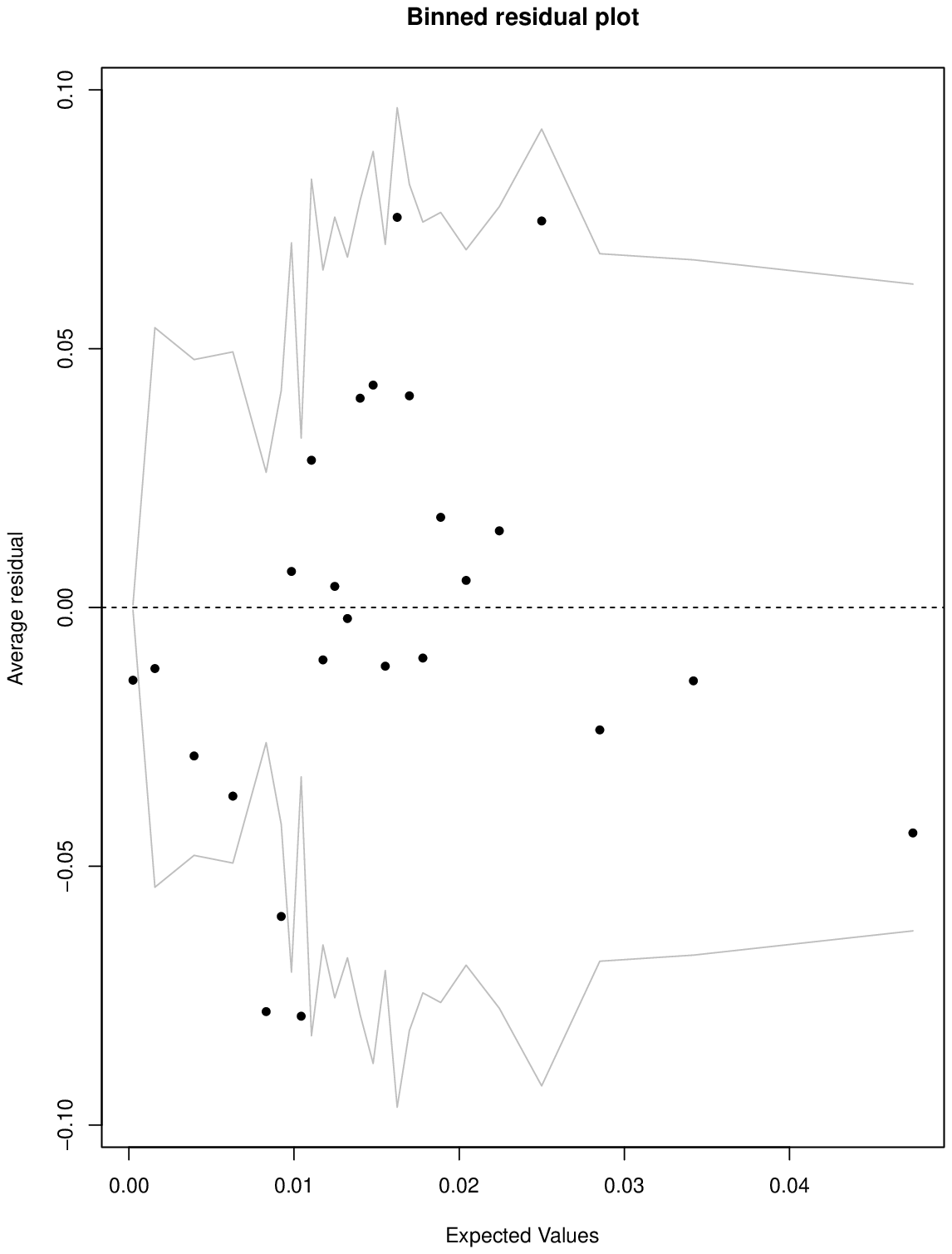}
}
\caption{Average Pearson residual against average fitted value for $25$ bins for model
(\ref{e:LGCP-simple}). The grey lines correspond to two standard deviations bounds.}
\label{F:binnedplot}
\end{figure}

We also need to validate our model. To do so, consider the Pearson residuals
\[
   \frac{ n(s_i, t_j) - \lambda(s_i, t_j; \hat \zeta) \Delta \Delta(s_i) }
  { \sqrt{
     \hat \lambda(s_i, t_j; \hat \zeta)  \Delta \Delta(s_i)
     +
     \lambda(s_i, t_j; \hat \zeta)^2 ( e^{4\hat \alpha^2\hat \sigma^2} - 1 ) \Delta^2 \Delta(s_i)^2
    }
}
\]
for $t_j > t_0$ and all $s_i$. Because the observed incidences take only very small values, 
residual plots are not helpful in assessing the model fit. A better alternative is to divide the 
data into bins based on their fitted values and plot the average residuals against the average 
fitted value for each bin, as shown in Figure~\ref{F:binnedplot}. The fit seems to be adequate 
for most bins.

To summarise, after excluding $\eta$, our final model is a log-Gaussian Cox 
process with driving random measure
\begin{equation} \label{e:LGCP-simple}
\Delta \Delta(s_i) \exp\left[
\theta_1 + \theta_2 V(s_i,t_j) +
 \alpha \left( X(s_i, t_0) - X(s_i, t_j) \right)  
\right].
\end{equation}
It is interesting to observe that a purely temporal analysis \cite{BakiLies22} yielded a
similar model with a time component instead of a drop in pore pressure.

\section{Monitoring seismic hazard}
\label{S:monitor}

Monitoring is based on the posterior distribution of $\Lambda$ or, 
equivalently, $E$ given the recorded earthquakes. Write 
$\nn = (n(s_i, t_j))_{i, j}$ for the vector of observed earthquake
counts $n(s_i, t_j)$  in the cells indexed by the $s_i$ and $t_j$. Then,
upon ignoring all terms that do not depend on $E$,
the log posterior likelihood reads
\begin{equation}
\log f( e(s_i, t_j)_{i,j} | \nn)  =  
- \sum_{(s_i, t_j)} \frac{ e(s_i, t_j)^2}{2 \sigma^2}   +
 \sum_{(s_i,t_j)} \left\{ n(s_i, t_j) \log \Lambda_{e}( s_i, t_j )
 - \Delta \Delta_S \Lambda_{e}( s_i, t_j )
\right\}
\label{e:posteriorE}
\end{equation}
where the $s_i$ range through the cell representatives in $W_S$ and
$t_j$ indicate the time intervals. We write $\Lambda_e$ to emphasise 
the dependence of (\ref{e:driving}) on the realisation $e$ of $E$. 
The marginal posterior likelihood for fixed $s_i$ will be denoted by 
$f_{s_i}( e(s_i, t_j)_j | \nn)$. 

We use a Markov chain Monte Carlo technique, the Metropolis adjusted
Langevin algorithm (MALA) proposed by Julian Besag \cite{Besa94},
to draw samples from (\ref{e:posteriorE}). The algorithm is a 
Metropolis--Hastings sampler \cite{MengTwee96} in which moves
are proposed in the direction of the gradient of (\ref{e:posteriorE}).
It is important to observe that the independence of our model across the 
spatial domain allows for parallel implementation. Thus, for each spatial 
grid cell around $s$, we run the following algorithm for $e(s, t_0), \dots,
e(s, t_m)$.

\begin{algo} \mbox{} If the current state is $\ee(s) = (e(s,t_0), \dots, 
e(s, t_m))$ and the earthquake count vector is $\nn$, then\\
\begin{enumerate}
\item sample a realisation $\tilde e(s,t_0), \dots, \tilde e(s, t_m)$ from 
independent normal distributions with variance $h$ and mean 
\[
\mu(s, t_j; e) = \left(1 - \frac{h}{2\sigma^2}\right) e(s, t_j)
- \frac{h}{2}  
 \alpha \left\{ n(s, t_j) - \Lambda_e(s, t_j) \Delta \Delta(s)
\right\}
\]
for $j>0$ and
\[
\mu(s, t_0; e) = \left(1 - \frac{h}{2\sigma^2}\right) e(s, t_0)
+ \frac{h}{2} \alpha 
 \sum_{i = 1}^{m} \left\{
 n(s, t_i)  - \Lambda_e(s, t_i) \Delta \Delta(s)
\right\};
\]
\item accept the new state with probability
\[
\frac{ 
  f_s(\tilde e(s, t_0), \dots, \tilde e(s, t_m) | \nn) 
\exp( - \sum_{j=0}^m ( e(s, t_j) -  \mu(s, t_j; \tilde e))^2 / (2h)  ) }
{
 f_s(e(s, t_0), \dots, e(s, t_m) | \nn)
 \exp( - \sum_{j=0}^m ( \tilde e(s, t_j) -  \mu(s, t_j; e))^2 / (2h)  ) }.
\]
\end{enumerate}
\end{algo}

Since the proposals are governed by a normal distribution, which has a strictly positive 
probability density, by \cite[Lemma~1.1]{MengTwee96}, the Markov chain generated by the
MALA algorithm above is $f_s$-irreducible. Also $f_s$ is a strictly positive probability 
density on $\oR^m$, so by \cite[Lemma~1.2]{MengTwee96} the Markov chain is also aperiodic.
By construction, $f_s$ is an invariant measure. We conclude that the Markov chain converges
in total variation from almost all initial states \cite[Proposition~7.7]{MollWaag04}.

\begin{figure}[htb]
\vspace{-0.5in}
\centering
\centerline{
\epsfxsize=0.95\hsize
\epsffile{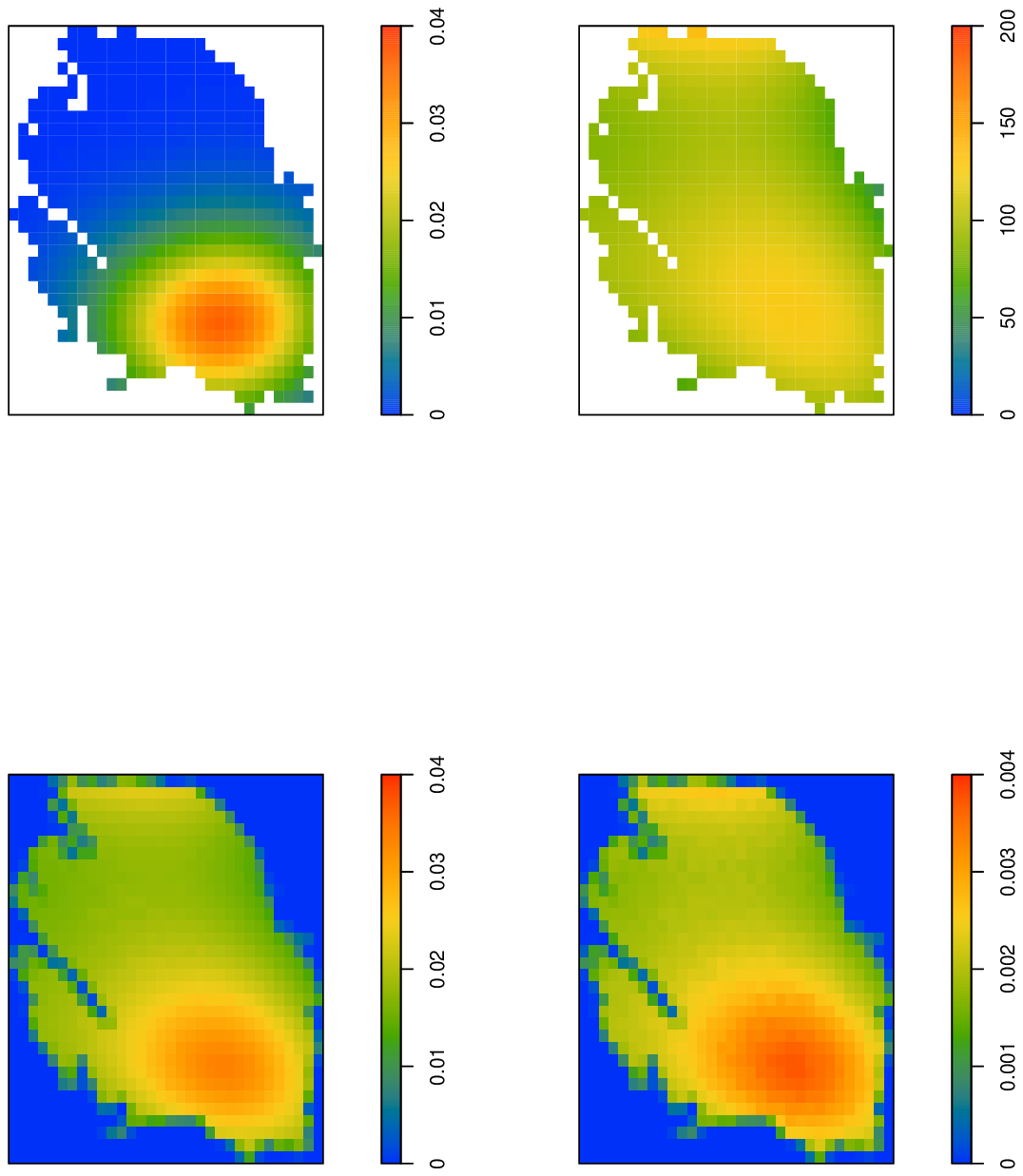}
}
\caption{Top left: Smoothed gas production over 2021 (in Nbcm for each grid cell).
Top right: Estimated pressure drop (in bara for each grid cell) from 1995 until 2022.
Bottom left: Mean posterior number of earthquakes in 2022 (for each grid cell, sample
size $I=5,000$).
Bottom right: Sample standard deviation of posterior number of earthquakes in 2022 
(for each grid cell, sample size $I=5,000$).}
\label{F:posterior}
\end{figure}

Having obtained samples from the posterior distribution of $\Lambda$, we are ready to 
monitor earthquake hazard.  Monitoring targets the posterior distribution of counts in 
the time interval $[t_{m+1}, t_{m+1} + \Delta]$, that is, in the year 2022. For our 
Groningen data, recalling (\ref{e:LGCP-simple}), these counts follow a Poisson distribution 
with intensity 
\[
\exp\left[ \hat \theta_0 + \hat \theta_1 V(s, t_m + \Delta) +
\hat \alpha \left\{ X_i (s, t_0 ) - m(s, t_m + \Delta) - E_i(s, t_m + \Delta) \right\}
\right] \times \Delta \Delta(s)
\]
where the family $\left\{ X_i(s,t_0) = m(s, t_0) + E_i(s, t_0) \right\}_{i=1, \dots, I}$ 
indexed by cell representatives $s$ use the samples $E_i(s,t_0)$
from the posterior given the counts generated by the MALA algorithm above, and 
$E_i(s, t_m + \Delta )$ is white noise with variance $\hat \sigma^2$.

\begin{figure}[thb]
\centering
\centerline{
\epsfxsize=0.45\hsize
\epsffile{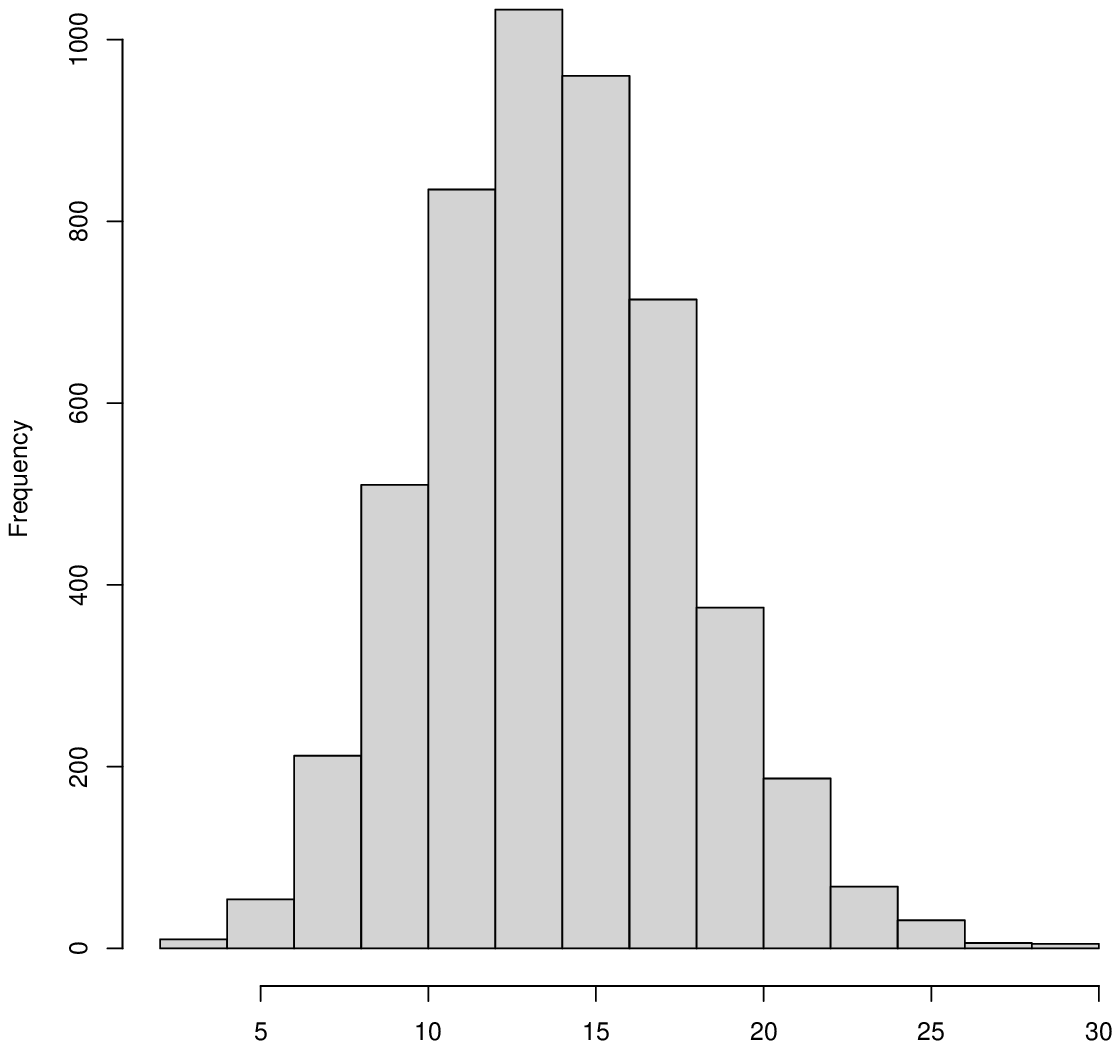}
}
\caption{Histogram of a sample of size $I=5,000$ from the posterior distribution of the number 
of earthquakes in 2022.}
\label{F:histcounts}
\end{figure}

With $h=0.02 \hat \sigma$, a burn-in of $10,000$ steps for each spatial grid cell
and subsampling every $1,000$ steps, we plot the mean and standard deviation of 
$\Lambda(s,t_m+\Delta) \Delta \Delta(s)$, intensity of earthquakes in each spatial 
cell around $s$ for the year 2022, in the bottom row of Figure~\ref{F:posterior}. 
The top row in the same figure depicts the covariates, namely the gas production 
figures in the preceding year and the estimated drop in pressure up to 2022. We show 
the histogram of earthquake counts for a sample of size $I = 5,000$ in 
Figure~\ref{F:histcounts}. 

The total volume of gas extracted in 2021 was low, $6.48$ Nbcm compared to around
over $50$ Nbcm in 2013, and concentrated in the south of the gas field. As for the 
pressure, it can be seen that the estimated decrease in pressure is smallest in the 
western and eastern periphery. Because the wells in the south were taken into production 
earlier than those in the north, initially a larger drop in pressure was measured 
in the south. To reduce this imbalance somewhat, in the seventies, eighties and 
nineties, the northern locations were preferred for production. However, in response 
to concerns following a major earthquake, the Dutch government imposed production 
caps on some northern clusters in 2014, which again emphasised the larger drop in pore 
pressure in the south. Additionally, the top-right plot in Figure~\ref{F:posterior} 
shows a larger estimated fall in pressure in the far north offshore part of the field. 
Indeed, the observation well at Oldorp in the north-western corner of the field is 
known to be atypical for the field: quite high values were observed in 1995 and 
there are very few recent measurements. We refer to \cite{JageViss17} for a more 
detailed description of the geology of the Groningen field. 

The risk map in the lower left panel of Figure~\ref{F:posterior} reflects these features. 
The area of increased risk due to gas extraction in the south of the field is tilted 
according to the pore pressure gradient, there is a smaller risk in the peripherical 
regions and the absence of production in the far north offsets the large drop in pressure. 
The standard deviation of the posterior is highest in the large production region in the 
southeast and in the offshore northern region. 

We also plot the histogram of the posterior earthquake count over 2022. For comparison, 
the actual number of earthquakes was $12$. 

\clearpage

\section{Conclusion}
\label{S:conclude}

In this paper, we explored the seismic risk in the Groningen gas field by modifying
the  state-of-the-art rate-state model in two directions, namely allowing for noise in 
pore pressure measurements and explicitly taking into account gas production volumes.
We investigated the first- and second-moment measures of the resulting Cox process 
and estimated its parameters by a taylor-made estimating equation. We then constructed 
a Markov Chain Monte-Carlo algorithm to monitor seismicity.

An important feature of our approach is that it is completely data-driven and does not
rely on reservoir models. The advantage of working in a data-driven fashion is that 
erroneous model assumptions cannot impact the monitoring and that uncertainty 
quantification is part of the toolbox; a drawback is that the accuracy depends on 
the quality of the data at hand. In our context, from the mid 1990s, the earthquake 
catalogue maintained by the Royal Netherlands Meteorological Office (KNMI) is accurate. Production 
figures too are available at various public websites, but not always accurate. Therefore
we used the data provided directly by the production company NAM for best results. 
Pore pressure measurements, however, are available only at wells and are quite sparse. 
Especially in recent years, there are not many observations. In view of the facts
that caps on production were put in place in 2014 and that production stopped altogether 
in 2024, which will likely affect the future drop in pore pressure, we recommend 
stepping up efforts in pore pressure measurements to improve the monitoring of seismic hazard.

The model can be extended in various directions. For instance, other explanatory variables, 
such as information on fault lines and subsidence, or other geological features of the field, 
could be taken into account. 
Also, spatially correlated random factors could be added to the model. From a theoretical 
point of view, an asymptotic theory for the estimating equation would be of interest. Finally, 
we intend to compare our data-driven approach with one that inputs the NAM reservoir model 
for the pore pressure.

\section*{Acknowledgements}

This research was funded by the Dutch Research Council NWO through
their DEEPNL programme (grant number DEEP.NL.2018.033).
We are grateful to Mr Van Eijs for providing us with data on gas production.

\newpage

\section*{Appendix A: Calculation of $\oE\Gamma(s, t_k)$ and $\cov( \Gamma(s, t_k), \Gamma(s,t_l))$}
\label{A:ProofState}

Since
\[
\oE e^{t Z} = \exp\left( \mu t + \sigma^2 t^2 / 2 \right)
\]
when $Z$ is normally distributed with mean $\mu$ and variance $\sigma^2$,
the formula for $\oE\Gamma(s, t_k)$ in Section~\ref{S:state} follows immediately. 
Moreover, for $0 < k \leq l$,
\begin{eqnarray}
\nonumber
\cov( \Gamma(s, t_k), \Gamma(s, t_l) ) & = &
\alpha^2 \Delta^2
\sum_{i=0}^{k-1} \sum_{j=0}^{l-1} \cov\left( 
e^{\alpha \left[ X(s, t_k ) - X(s, t_i ) \right]}, 
e^{\alpha \left[ X(s, t_l ) - X(s, t_j ) \right]} 
\right) \\ 
\nonumber
& + &
\alpha \Delta \gamma_0 \sum_{i=0}^{k-1}
\cov\left(
e^{\alpha \left[ X(s, t_k ) - X(s, t_i ) \right]}, 
e^{\alpha \left[ X(s, t_l ) - X(s, t_0) \right]} 
\right)  \\
\nonumber
& + &
 \alpha \Delta \gamma_0 \sum_{j=0}^{l-1}
\cov\left(
e^{\alpha \left[ X(s, t_k ) - X(s, t_0) \right]}, 
e^{\alpha \left[ X(s, t_l ) - X(s, t_j) \right]} 
\right)  \\
& + &
\label{e:covterms}
\gamma_0^2 \cov\left(
e^{\alpha \left[ X(s, t_k ) - X(s, t_0) \right]}, 
e^{\alpha \left[ X(s, t_l ) - X(s, t_0) \right]} 
\right).
\end{eqnarray}
Let us work out the terms in the expression on the right-hand side 
of equation (\ref{e:covterms}) one by one. First, consider the double sum 
over $i$ and $j$. Because of the independence of the components of 
the random vector $E$, summands for which $k$, $l$, $i$ and $j$ 
are all different do not contribute. Therefore, for $0 < k = l$, 
the total contribution of the double sum is
\[
\alpha^2 \Delta^2 c^2 \sum_{i=0}^{k-1} f_{ki}^2 \left( c^2 - 1 \right) +
\alpha^2 \Delta^2 c^2 \sum_{i=0}^{k-1} f_{ki} 
     \sum_{j\neq i; j=0}^{k-1} f_{kj} \left( c - 1 \right).
\]
For $0 < k < l$, the double sum contributes non-zero entries for $i=j$ 
(which cannot be equal to $k$ or $l$) and for $i\neq j = k$.
Their total contribution is
\[
\alpha^2 \Delta^2 c^2 \sum_{i=0}^{k-1} f_{ki} f_{li}
 \left( c - 1 \right) 
- \alpha^2 \Delta^2 c^2 \sum_{i=0}^{k-1} f_{li} \left(1 - \frac{1}{c} \right).
\]
Next consider the two single sums on the right-hand side of (\ref{e:covterms}).
If $0 < k = l$, they are identical and each one is equal to
\[
\alpha \Delta \gamma_0 c^2 \left[
 f_{k0}^2 (c^2 - 1)  + (c-1) \sum_{i=1}^{k-1} f_{ki} f_{k0}
\right].
\]
In the case that $0 < k< l$, the sum over $i$ has a non-zero contribution
only for $i=0$, and the sum over $j$ has non-vanishing contributions for 
$j=0$ and $j=k$. Adding these up, we obtain
\[
\alpha \Delta \gamma_0 c^2 f_{l0} \left[
   2 f_{k0} (c - 1) - \left( 1 - \frac{1}{c} \right)
\right].
\]
Finally, the last term in the expression on the right-hand side 
of equation (\ref{e:covterms}) 
reads $\gamma_0^2 f_{k0}^2 $  $c^2 ( c^2 - 1 )$ for $0 < k = l$ and 
$\gamma_0^2 f_{k0}f_{l0} c^2 (c - 1 )$ when $0 < k < l$.
Expression (\ref{e:covsum}) now follows from tallying up the 
various contributions. The variance of the state variable is obtained
by taking $k=l$.

\bigskip

\noindent
When the pore pressure increases, the state variables are 
non-negatively correlated. To see this, consider $0 \leq i < k < l$. 
Then
\[
\alpha^2 \Delta^2  f_{li} \{ f_{ki} c^2 (c-1) - c ( c-1) \} \geq 0
\]
if $\sigma^2 = 0$ or $c f_{ki} \geq 1$. The latter condition 
is equivalent to 
\[
 \alpha^2 \sigma^2 + \alpha \{ m(s, t_k) - m(s, t_i) \} \geq 0
\]
and is implied by the assumption that $m$ is increasing. By the 
same argument, for increasing pore pressure,
$\alpha \Delta \gamma_0 f_{l0} \{ f_{k0} c^2 (c-1) - c (c-1) \} \geq 0$
and an appeal to (\ref{e:covsum}) completes the proof.
 
\bigskip

\noindent
Next, suppose that the pore pressure decreases.
For $0 \leq i < k < l$, consider
\(
\alpha^2 \Delta^2  f_{li} c (c-1) \{ c f_{ki}  - 1 \} 
\).
The term in between curly brackets is negative if and only if 
\[
 \alpha \left\{ m(s, t_k ) - m(s, t_i) \right\} + \alpha^2 \sigma^2  < 0.
\]
Since $m$ is decreasing, if $\alpha \sigma^2 < 
\min_i \left\{ m(s, t_i)  - m(s, t_{i+1)}) \right\}$,
\[
 \alpha^2 \sigma^2 < m(s, t_i) - m(s, t_{i+1}) \leq m(s, t_i) - m(s, t_k)
\]
so that
\(
\alpha^2 \Delta^2  f_{li} c (c-1) \{ c f_{ki}  - 1 \} \leq 0
\).
Since the same argument can be used to show that 
\(
\alpha \Delta \gamma_0  f_{l0} c (c-1) \{ c f_{k0}  - 1 \} \leq 0
\),
\[
\cov( \Gamma(s, t_k), \Gamma(t, t_l) )
 - (\alpha \Delta \gamma_0 + \gamma_0^2) c^2 (c-1) f_{k0} f_{l0} \leq 0.
\]
If $\alpha \sigma^2 > m(0)$, then
\[
 \alpha^2 \sigma^2 > m(s, t_0) \geq m(s, t_i) \geq m(s, t_i) - m(s, t_k)
\]
and therefore 
\(
\alpha^2 \Delta^2  f_{li} c (c-1) \{ c f_{ki}  - 1 \} \geq 0
\).
Similarly,
\(
\alpha \Delta \gamma_0  f_{l0} c (c-1) \{ c f_{k0}  - 1 \} \geq 0.
\)
Consequently, $\cov( \Gamma(s, t_k), \Gamma(t, t_l) ) \geq 0$.

\newpage

\section*{Appendix B: Approximate momens of the rate variable}
\label{A:AppendixRate}

To approximate the expectation of $1/\Gamma(s, t_k)$, 
apply the delta method based on the Taylor expansion
\[
\frac{1}{x_0+h} \approx \frac{1}{x_0} - \frac{h}{x_0^2} + \frac{1}{2!}
\frac{ 2h^2}{x_0^3}
\]
around $x_0$ equal to the expectation of $\Gamma(s, t_k)$. 
Upon taking the expectation, one obtains that
\[
 \oE \left[ \frac{1}{\Gamma(s, t_k)} \right]
 \approx  
\frac{1}{\oE \Gamma(s, t_k)} 
- \oE \left[ 
  \frac{\Gamma(s, t_k) - \oE \Gamma(s, t_k)}{
(\oE \Gamma(s, t_k) )^2 }  \right] 
 +  \oE \left[ \frac{ ( \Gamma(s, t_k ) - \oE\Gamma(s, t_k) )^2}
{ (\oE \Gamma(s, t_k))^3} \right].
\]
The middle term on the right-hand side is zero, and (\ref{e:mean}) follows.

\bigskip

\noindent
The Taylor expansion of the function $(x,y) \mapsto 1/(xy)$ around the 
point $(\oE\Gamma(s, t_k),$ $ \oE\Gamma(s, t_l) )$ yields the approximation
\begin{eqnarray*}
 \frac{1}{\Gamma(s, t_k)} 
 \frac{1}{\Gamma(s, t_l)} 
& \approx &
\frac{1}{\oE \Gamma(s, t_k)} 
\frac{1}{\oE \Gamma(s, t_l)}  
+
 \frac{ 
\left(
 \Gamma(s, t_k) - \oE \Gamma(s, t_k)
\right) \,
\left(
 \Gamma(s, t_l) - \oE \Gamma(s, t_l)
\right)
}{ (\oE \Gamma(s, t_k) )^2 ( \oE \Gamma(s, t_l) )^2} 
\\
& - &
  \frac{\Gamma(s, t_k) - \oE \Gamma(s, t_k)}{
(\oE \Gamma(s, t_k) )^2  \oE \Gamma(s, t_l)} 
-
 \frac{\Gamma(s, t_l) - \oE \Gamma(s, t_l)}{
(\oE \Gamma(s, t_l) )^2  \oE \Gamma(s, t_k)} 
\\
& +  &
 \frac{ 
\left(
 \Gamma(s, t_k) - \oE \Gamma(s, t_k)
\right)^2
}{ (\oE \Gamma(s, t_k) )^3 \oE \Gamma(s, t_l)} 
+
 \frac{ 
\left(
 \Gamma(s, t_l) - \oE \Gamma(s, t_l)
\right)^2
}{\oE \Gamma(s, t_k) ( \oE \Gamma(s, t_l) )^3} 
\end{eqnarray*}
up to second-order moments. Finally take  expectations to
obtain an approximation of the expected cross product of the rate.

\newpage

\section*{Appendix C: Partial derivatives for parameter estimation}
\label{A:Jacobians}

Write, for  $s\in W_S$ the centre of a cell with area $\Delta(s)$
and $t_k = t_0 + k \Delta$,
\[
S(s,t_k) = e^\eta \Delta \sum_{i=0}^{k-1}
  e^{-\alpha ( m(s, t_i) - m(s, t_k) ) }
+ 
  e^{-\alpha ( m(s, t_0) - m(s, t_k) ) }
\]
and let $h$ be as in Section~\ref{S:params}. Then 
\(
\frac{\partial}{\partial \theta_1} h(s, t_k) / { h(s, t_k) } =  1 
\)
and
\[
\frac{ \frac{\partial}{\partial \theta_2} h(s, t_k) }{ h(s, t_k) }
 =  V(s, t_k), \quad
\frac{
\frac{\partial}{\partial \alpha} h(s, t_k) }{ h(s, t_k) } 
 = - \frac{ \frac{\partial}{\partial \alpha} S(s, t_k)} {S(s, t_k)} , \quad
  \frac{ \frac{\partial}{\partial \eta} h(s, t_k) }{ h(s, t_k) }  
 =  
- \frac{ \frac{\partial}{\partial \eta} S(s, t_k)} {S(s, t_k)}.
\]
These derivatives define $F_\zeta$ in the estimating equation (\ref{e:Rasmus-modified}).
The partial derivatives of $S$ can be calculated recursively:
\begin{eqnarray*}
S(s,t_k) & = & \left( S(s,t_{k-1}) + e^\eta\Delta \right) e^{-\alpha(m(s,t_{k-1})-m(s,t_k))} \\
\frac{\partial}{\partial \alpha} S(s,t_k) & = & 
 \left( \frac{\partial}{\partial \alpha} S(s,t_{k-1}) \right)
   e^{-\alpha (m(s,t_{k-1})-m(s,t_k))} -  (m(s,t_{k-1})-m(s,t_k) ) S(s,t_k) \\
 \frac{\partial}{\partial \eta} S(s,t_k) & = & 
 \left( \frac{\partial}{\partial \eta} S(s,t_{k-1}) + e^\eta \Delta \right)
   e^{-\alpha (m(s,t_{k-1})-m(s,t_k))} \\
 \frac{\partial^2}{\partial \alpha^2} S(s,t_k) & = & 
 \left(  \frac{\partial^2}{\partial \alpha^2} S(s,t_{k-1}) \right)
   e^{-\alpha (m(s,t_{k-1})-m(s,t_k))} \\
& - & 
   \left(m(s,t_{k-1})-m(s,t_k) \right) \left[ 
     \frac{\partial}{\partial \alpha} S(s,t_k) 
     + \left(\frac{\partial}{\partial \alpha} S(s,t_{k-1}) \right)
    e^{-\alpha (m(s,t_{k-1})-m(s,t_k))} \right] \\
 \frac{\partial^2}{\partial\alpha \partial \eta} S(s,t_k) & = & 
\left( \frac{\partial^2}{\partial\alpha \partial \eta} S(s,t_{k-1}) \right)
    e^{-\alpha (m(s,t_{k-1})-m(s,t_k))} -  
   \left (m(s,t_{k-1})-m(s,t_k) \right) \frac{\partial}{\partial \eta}S(s,t_k)
\end{eqnarray*}
and $\frac{\partial^2}{\partial \eta^2} S(s,t_k)  =  \frac{\partial}{\partial \eta} S(s,t_k)$.

\noindent
Equation (\ref{e:Rasmus-modified}) may be solved  by the 
Newton method that iteratively solves the equation
\[
J_F(\zeta_n) ( \zeta_{n+1} - \zeta_n ) = - F(\zeta_n)
\]
for $\zeta_{n+1} - \zeta_n$.
The Jacobian $J_F(\zeta)$ is a four-by-four matrix with components
\begin{equation*}
\begin{array}{l|l}
J_F(\zeta)_{1,1} = - \sum_{(s_i,t_j)} \widehat \lambda(s_i,t_j) \Delta \Delta(s_i) &
J_F(\zeta)_{2,1} = - \sum_{(s_i,t_j)} V(s_i,t_j) \widehat \lambda(s_i,t_j) \Delta \Delta(s_i)
\\
J_F(\zeta)_{1,2} = - \sum_{(s_i,t_j)} V(s_i,t_j) \widehat \lambda(s_i,t_j) \Delta \Delta(s_i) &
J_F(\zeta)_{2,2} = - \sum_{(s_i,t_j)} V(s_i,t_j)^2 \, \widehat \lambda(s_i,t_j) \Delta \Delta(s_i)
\\ 
J_F(\zeta)_{1,3} =  - \sum_{(s_i,t_j)} \frac{\partial \widehat \lambda(s_i,t_j)}{\partial\alpha} \Delta \Delta(s_i) &
J_F(\zeta)_{2,3} = - \sum_{(s_i,t_j)} V(s_i,t_j) \frac{\partial\widehat \lambda(s_i,t_j)}{\partial\alpha}
   \Delta \Delta(s_i) 
\\ 
J_F(\zeta)_{1,4} = - \sum_{(s_i,t_j)} \frac{\partial \widehat \lambda(s_i,t_j)}{\partial \eta} 
      \Delta \Delta(s_i)&
J_F(\zeta)_{2,4} = - \sum_{(s_i,t_j)} V(s_i,t_j) \frac{\partial\widehat \lambda(s_i,t_j)}{\partial\eta}
   \Delta \Delta(s_i)\\
& \\
J_F(\zeta)_{3,1} =  \sum_{(s_i,t_j)} \frac{\frac{\partial S(s_i,t_j)}{\partial\alpha}}{S(s_i,t_j)}
  \widehat \lambda(s_i,t_j) \Delta \Delta(s_i) & 
J_F(\zeta)_{3,2} = \sum_{(s_i,t_j)} \frac{\frac{\partial S(s_i,t_j)}{\partial\alpha}}{S(s_i,t_j)}
  V(s_i,t_j) \widehat \lambda(s_i,t_j) \Delta \Delta(s_i) \\ 
J_F(\zeta)_{4,1} =  \sum_{(s_i,t_j)} \frac{\frac{\partial S(s_i,t_j)}{\partial\eta}}{S(s_i,t_j)}
  \widehat \lambda(s_i,t_j) \Delta \Delta(s_i) & 
J_F(\zeta)_{4,2} =  \sum_{(s_i,t_j)} \frac{\frac{\partial S(s_i,t_j)}{\partial\eta}}{S(s_i,t_j)}
  V(s_i,t_j) \widehat \lambda(s_i,t_j) \Delta \Delta(s_i).
\end{array}
\end{equation*}
In none of these the counts play a role. They do for the last two terms in 
the third row:
\[
\sum_{(s_i,t_j)} \left\{
 \left(\frac{ \frac{\partial S(s_i,t_j)}{\partial \alpha} }{S(s_i,t_j)}\right)^2
- \frac{ \frac{\partial^2 S(s_i,t_j)}{\partial \alpha^2} }{S(s_i,t_j)}
\right\} \left[ N(s_i,t_j) - \widehat \lambda(s_i,t_j) \Delta \Delta(s_i) \right]
+
\]
\[
\sum_{(s_i,t_j)} 
  \frac{ \frac{\partial S(s_i,t_j)}{\partial \alpha}}{S(s_i,t_j)}
\frac{\partial \widehat \lambda(s_i,t_j)}{\partial \alpha} \Delta \Delta(s_i)
\]
and
\[
\sum_{(s_i,t_j)} \left\{
   \frac{ \frac{\partial S(s_i,t_j)}{\partial \alpha} }{S(s_i,t_j)}
  \frac{ \frac{\partial S(s_i,t_j) }{\partial \eta} }{S(s_i,t_j)}
- \frac{ \frac{\partial^2 S(s_i,t_j)}{\partial \alpha\partial \eta} }{S(s_i,t_j)}
\right\} \left[ N(s_i,t_j) - \widehat \lambda(s_i,t_j) \Delta \Delta(s_i) \right]
+
\]
\[
\sum_{(s_i,t_j)} 
  \frac{ \frac{\partial S(s_i,t_j)}{\partial \alpha}}{S(s_i,t_j)}
\frac{\partial \widehat \lambda(s_i,t_j)}{\partial \eta} \Delta \Delta(s_i).
\]
Finally, the two last entries of the fourth row are
\[
\sum_{(s_i,t_j)} \left\{
   \frac{ \frac{\partial S(s_i,t_j)}{\partial \alpha} }{S(s_i,t_j)}
  \frac{ \frac{\partial S(s_i,t_j) }{\partial \eta} }{S(s_i,t_j)}
- \frac{ \frac{\partial^2 S(s_i,t_j)}{\partial \alpha\partial \eta} }{S(s_i,t_j)}
\right\} \left[ N(s_i,t_j) - \widehat \lambda(s_i,t_j) \Delta \Delta(s_i) \right]
+
\]
\[
\sum_{(s_i,t_j)} 
  \frac{ \frac{\partial S(s_i,t_j)}{\partial \eta}}{S(s_i,t_j)}
\frac{\partial \widehat \lambda(s_i,t_j)}{\partial \alpha} \Delta \Delta(s_i)
\]
and
\[
\sum_{(s_i,t_j)} \left\{
  \left( \frac{ \frac{\partial S(s_i,t_j) }{\partial \eta} }{S(s_i,t_j)}\right)^2
- \frac{ \frac{\partial^2 S(s_i,t_j)}{\partial \eta^2} }{S(s_i,t_j)}
\right\} \left[ N(s_i,t_j) - \widehat \lambda(s_i,t_j) \Delta \Delta(s_i) \right]
+
\]
\[
\sum_{(s_i,t_j)} 
  \frac{ \frac{\partial S(s_i,t_j)}{\partial \eta}}{S(s_i,t_j)}
\frac{\partial \widehat \lambda(s_i,t_j)}{\partial \eta} \Delta \Delta(s_i).
\]
Note that the Jacobian is not symmetric as it does not correspond to a likelihood!
A recursive formula for the partial derivatives of $S$ is given above.
It remains to calculate the  partial derivatives of $\widehat \lambda$.
Those with respect to $\theta_1$ and $\theta_2$ are, respectively, $\widehat \lambda$
and $V \widehat\lambda$. Furthermore,
\begin{align*}
& \frac{\partial \hat{\lambda}(s,t_k)}{\partial\alpha} =
   e^{\theta_1 + \theta_2 {V}(s,t_k)}
   \frac{1}{L} \sum_{l=1}^L \frac{-\frac{\partial}{\partial \alpha}
   S_{X_l}(s,t_k)}{S_{X_l}(s,t_k)^2}\\
& \frac{\partial \hat{\lambda}(s,t_k)}{\partial\eta} =  
   e^{\theta_1 + \theta_2 {V}(s,t_k)}
   \frac{1}{L}\sum_{l=1}^L \frac{-\frac{\partial}{\partial \eta}
   S_{X_l}(s,t_k)}{S_{X_l}(s,t_k)^2}
\end{align*}
with 
\begin{align*}
S_{X_l} (s, t_k)  & = e^\eta \Delta \sum_{i=0}^{k-1} e^{-\alpha ( {X_l} (s,t_i) - {X_l} (s, t_k))} +
                   e^{-\alpha ( {X_l} (s,t_0) - {X_l} (s,t_k) ) } \\
 \frac{\partial}{\partial\alpha}S_{X_l}(s,t_k) & = 
      - (X_l(s, t_0) - X_l(s,t_k)) e^{- \alpha (X_l(s, t_0) - X_l(s,t_k))} \\
& - e^{\eta}\Delta\sum_{i=0}^{k-1} (X_l(s,t_i) - X_l(s,t_k)) e^{-\alpha (X_l(s,t_i) - X_l(s,t_k))}\\
        \frac{\partial}{\partial\eta}S_{X_l}(s,t_k) & = 
   e^{\eta}\Delta\sum_{i=0}^{k-1} e^{-\alpha (X_l(s,t_i) - X_l(s,t_k))}.
\end{align*}

\newpage
\section*{Appendix D: Godambe matrix}

The Godambe matrix of the estimating equation (\ref{e:Rasmus-modified}) 
with $e^\eta = 0$ and remaining parameter vector
$\zeta = ( \theta_1, \theta_2, \alpha)$ 
is of the form $U \Sigma_F^{-1} U$. In the limit upon letting $\Delta$ 
and all $\Delta(s)$ go to zero, the first two columns of $U$ read
\begin{equation*}
\left[ 
\begin{array}{ll}
 \int_{W_S\times W_T} \lambda(s,t; \zeta) ds dt &
 \int_{W_S\times W_T} V(s,t) \lambda(s,t; \zeta) ds dt 
 \\
 \int_{W_S\times W_T} V(s,t) \lambda(s,t; \zeta) ds dt &
 \int_{W_S\times W_T} V(s,t)^2 \lambda(s,t; \zeta) ds dt 
 \\
 \int_{W_S\times W_T} ( m(s, t_0) - m(s,t) ) \lambda(s,t; \zeta) ds dt &
 \int_{W_S\times W_T} ( m(s, t_0) - m(s,t) ) V(s,t) \lambda(s,t; \zeta) ds dt 
\end{array}
\right] 
\end{equation*}
for 
\[
 \lambda(s,t; \zeta) =  \exp\left[ \theta_1 + \theta_2 V(s,t) +
 \alpha( m(s, t_0) - m(s,t) ) + \alpha^2 \sigma^2 \right].
\]
The last column of $U$ is
\begin{equation}
\left[ 
\begin{array}{l}
 \int_{W_S\times W_T} l(s,t; \zeta) ds dt \\
 \int_{W_S\times W_T} V(s,t) l(s,t; \zeta) ds dt  \\
 \int_{W_S\times W_T} ( m(s, t_0) - m(s,t) ) l(s,t; \zeta) ds dt
\end{array}
\right] 
\label{e:U}
\end{equation}
where 
\[
l(s,t; \zeta) = \oE_{\zeta} \left\{ (X(s, t_0) - X(s,t)) e^{ \theta_1 + \theta_2 V(s,t) }
e^{\alpha ( X(s,t_0) - X(s,t) )} \right\}.
\]

\bigskip

\noindent
The first two columns of the matrix $\Sigma_F$ are identical to those of $U$. Its last
column reads
\begin{equation}
\left[ 
\begin{array}{l}
 \int_{W_S\times W_T} ( m(s, t_0) - m(s,t) ) \lambda(s,t; \zeta) ds dt \\
 \int_{W_S\times W_T} ( m(s, t_0) - m(s,t) ) V(s,t) \lambda(s,t; \zeta) ds dt \\
 \int_{W_S\times W_T} ( m(s, t_0) - m(s,t) )^2 \lambda(s,t; \zeta) ds dt 
\end{array}
\right].
\label{e:V}
\end{equation}

\end{document}